\documentclass[12pt]{article}
\usepackage{indentfirst}
\usepackage{bm}
\usepackage{graphics,color}
\usepackage{graphicx}
\usepackage{amsmath}
\usepackage{cases}
\allowdisplaybreaks
 \title{Protecting and enhancing spin squeezing under decoherence using weak measurement}
 
\begin{document}
\date{}
\maketitle \baselineskip=0.82cm
\author{Xiang-Ping Liao\footnote {Corresponding author. Tel:+86-731-22183457; Fax:+86-731-22183457. E-mail address: liaoxp1@126.com} $\;\;\;\;\;\;\;\;\;\;\;\;\;\;\;\;\;$Man-Sheng Rong$^{1}$$\;\;\;\;\;\;\;\;\;\;\;\;\;\;\;\;\;\;\;\;\;\;\;\;\;\;\;\;$ Mao-Fa Fang$^{2}$}

\begin{center}
$^{1}$ College of Science, Hunan University of Technology, Zhuzhou, Hunan 412008, China\\
$^{2}$ College of Physics and Information Science, Hunan Normal
University, Changsha, Hunan 410081, China\\

\end{center}
\vspace{1.5cm}

{\bf Abstract}

We propose an efficient method to protect spin squeezing under the action of amplitude-damping, depolarizing and phase-damping channels  based on measurement reversal from weak measurement, and consider an ensemble of N independent
spin-1/2 particles with exchange symmetry. We find that spin squeezing can be enhanced greatly under three different decoherence channels and spin-squeezing sudden death (SSSD) can be avoided undergoing amplitude damping and phase-damping channels.


{\bf PACS number(s)}:  03.65.Ud, 03.67.Mn, 03.65.Yz

\section{Introduction}

Spin squeezing has attracted a lot of
attention in both the theoretical and experimental fields for
decades$^{\cite{a1,a2,a3,a4,a5,a6,a7,a8}}$. An important application of spin squeezing
is to detect quantum entanglement$^{\cite{a9,a10,a11}}$. Due to the fact that spin squeezing is relatively
easy to be generated and measured$^{\cite{a2,a12,a13,a14}}$, spin-squeezing parameters are
multipartite entanglement witness in a general sense.
Lots of efforts have been devoted to find relations between
spin squeezing and entanglement$^{\cite{a4,a5,a6,a7,a15,a16,a17}}$. Another
application of spin squeezing is to improve the precision
of measurements such as leading-noise reduction $^{\cite{a18}}$ and improving
atomic sensor precision $^{\cite{a19}}$. Thus, spin-squeezed states are useful resources for quantum
information processing. However, the interactions between the system and its environment
usually cause decoherence. In practice, decoherence
is inevitable and harmful to spin squeezing and entanglement
$^{\cite{a23,a24,a25,a26,a27,a28,a29}}$.

We find that, analogous to the definition of entanglement sudden death (ESD) $^{\cite{a30}}$ and distillability sudden death(DSD)$^{\cite{a31}}$, spin squeezing can also
suddenly vanish with different lifetimes for some decoherence channels, showing in general different vanishing
times in multipartite correlations in quantum many-body
systems. Wang et al. $^{\cite{a28}}$  have found that, under local
decoherence, spin squeezing also appears as ¡°sudden death¡± similar to the discovery of pairwise entanglement sudden death. An method to protecting and enhancing spin squeezing via continuous dynamical decoupling is proposed by Adam Zaman Chaudhry et. al$^{\cite{a32}}$.

In 1988, weak measurement was introduced by Aharonov, Albert, and Vaidman (AAV)$^{\cite{a33}}$. Weak measurement is very useful and can
help understand many counterintuitive quantum phenomena, for example, Hardy's paradoxes $^{\cite{a34}}$. Recently, the weak measurement has been applied as a practically implementable method for protecting entanglement, quantum
fidelity of quantum states undergoing decoherence $^{\cite{a35,a36,a37,a38,a39,a40}}$ and improving payoffs in the quantum games in the presence of decoherence $^{\cite{a41}}$. However, the study on protecting spin squeezing under the action of decoherence and avoiding spin-squeezing sudden
death via using weak measurements is not involved so far.

 Motivated by recent studies
of decoherence effects on spin squeezing and the application of weak measurement, we propose an efficient method to avoid spin-squeezing sudden death
via measurement reversal from weak measurement, and consider an ensemble of N independent
spin-1/2 particles with exchange symmetry.

\section{The definitions of spin squeezing and concurrence}

We consider an ensemble of N spin-1/2 particles with
ground state $|1\rangle$ and excited state $|0\rangle$. This system has exchange
symmetry, and its dynamical properties can be described by
the collective operators
\begin{eqnarray}
 J_\alpha=\frac{1}{2}\sum_{k=1}^N \sigma_{k\alpha}
 \end{eqnarray}
 for $\alpha=x, y, z$. Here, $\sigma_{k\alpha}$ are the Pauli matrices for the kth
qubit.

We choose the initial state as a standard one-axis
twisted state$^{\cite{a1}}$
\begin{eqnarray}
 |\Psi(0)\rangle=e^{-i\theta J_{x}^2/2}|\downarrow...\downarrow\rangle
 \end{eqnarray}
 This state is prepared by the one-axis twisted Hamiltonian $H=\chi J_x^2$, with the coupling constant $\chi$ , and $\theta=2\chi t$ the twist
angle.

There are several spin-squeezing parameters, but we
list only three typical and related ones as follows$^{\cite{a1,a2,a3,a4,a5}}$:
\begin{eqnarray}
 \xi^2_{1}=\frac{4(\bigtriangleup J_{\vec{n}_{\bot}})^2_{min}}{N}
 \end{eqnarray}
 \begin{eqnarray}
 \xi^2_{2}=\frac{N^2}{4\langle \vec{J} \rangle^2}\xi^2_{1}
 \end{eqnarray}
 \begin{eqnarray}
 \xi^2_{3}=\frac{\lambda_{min}}{\langle \vec{J^2}\rangle-\frac{N}{2}}
 \end{eqnarray}
 Here, the minimization in the first equation is over all
directions denoted by $\vec{n}_{\perp}$, perpendicular to the mean spin
direction $\langle \vec{J} \rangle/\langle \vec{J^2} \rangle$; $\lambda_{min}$ is the minimum eigenvalue of the
matrix
\begin{eqnarray}
 \Gamma=(N-1)\gamma+\mathcal{C}
 \end{eqnarray}
where
\begin{eqnarray}
 \gamma_{kl}=C_{kl}-\langle J_{k}\rangle \langle J_{l}\rangle \;\;\; for\;\;\;   k, l\in\{x,y,z\}=\{1,2,3\}
 \end{eqnarray}
 is the covariance matrix and $\mathcal{C}=[C_{kl}]$ with
 \begin{eqnarray}
 C_{kl}=\frac{1}{2}\langle J_{l}J_{k}+J_{k}J_{l}\rangle
 \end{eqnarray}
 is the global correlation matrix. The parameters $\xi^2_{1}$, $\xi^2_{2}$ and $\xi^2_{3}$
were defined by Kitagawa and Ueda $^{\cite{a1}}$, Wineland et al. $^{\cite{a2}}$, and T\'{o}th et al. $^{\cite{a4}}$, respectively. If
 $\xi^2_{2}<1$ ($\xi^2_{3}<1$) spin squeezing occurs, and we can safely
say that the multipartite state is entangled.

For states with a well-defined parity (even or odd), we now express the squeezing
parameters in terms of local expectations and correlations$^{\cite{a7,a28}}$
\begin{eqnarray}
 \xi^2_{1}=1+2(N-1)(\langle \sigma_{1+}\sigma_{2-}\rangle-|\langle\sigma_{1-}\sigma_{2-}\rangle|)
 \end{eqnarray}
 \begin{eqnarray}
 \xi^2_{2}=\frac{\xi^2_{1}}{\langle \sigma_{1z}\rangle^2}
 \end{eqnarray}
  \begin{eqnarray}
  \xi^2_{3}=\frac{min\{\xi^2_{1},\varsigma^2\}}{(1-N^{-1})\langle\vec{\sigma}_1\cdot \vec{\sigma}_2\rangle+N^{-1}}
 \end{eqnarray}
 where
 \begin{eqnarray}
 \varsigma^2=1+(N-1)(\langle \sigma_{1z}\sigma_{2z}\rangle-\langle\sigma_{1z}\rangle \langle\sigma_{2z}\rangle)
 \end{eqnarray}
 For convenience, hereafter we use
 \begin{eqnarray}
 \zeta^2_k=max(0,1-\xi^2_k), k\in\{1,2,3\}
 \end{eqnarray}
 to characterize spin squeezing. With the above definition,
spin squeezing occurs when $\zeta^2_k >0$.

The concurrence is defined as $^{\cite{a45.5}}$
\begin{eqnarray}
C&=& max(0, \lambda_1-\lambda_2-\lambda_3-\lambda_4)
\end{eqnarray}
where $\lambda_i$  are the square roots of eigenvalues
of $ \tilde{\rho} \rho$. Here $\rho$ is the reduced density matrix of the
system, and
\begin{eqnarray}
\tilde{\rho}=(\sigma_y\otimes\sigma_y)\rho^*(\sigma_y\otimes\sigma_y)
\end{eqnarray}
where $\tilde{\rho}$ is the conjugate of $\rho$.

The two-spin reduced density matrix for a parity state
with the exchange symmetry can be written in a block-
diagonal form$^{\cite{a7}}$
\begin{eqnarray}
\rho_{12}=\left [
 \begin {array}{cccc}
 \upsilon_{+}&u^*\\
 u&\upsilon_{-}
 \end {array}
 \right ]
 \oplus
 \left [
 \begin {array}{cccc}
 w&y\\
 y&w
 \end {array}
 \right ]
 \end{eqnarray}
 where
 \begin{eqnarray}
\upsilon_{\pm}&=&\frac{1}{4}(1\pm2\langle\sigma_{1z}\rangle+\langle\sigma_{1z}\sigma_{2z}\rangle)\\\nonumber
w&=&\frac{1}{4}(1-\langle\sigma_{1z}\sigma_{2z}\rangle)\\\nonumber
u&=&\langle\sigma_{1+}\sigma_{2+}\rangle\\\nonumber
y&=&\langle\sigma_{1+}\sigma_{2-}\rangle
\end{eqnarray}
 the concurrence is given by
 \begin{eqnarray}
C&=& max\{0, 2(|u|-w),2(y-\sqrt{\upsilon_+\upsilon_-})\}
\end{eqnarray}

\section{Protecting spin squeezing under decoherence by using weak measurements}

We propose a scheme to protect spin squeezing under the action of decoherence channels by using weak measurement. The scheme is ¡°weak measurement M + decoherence channel + weak measurement N ¡±.

The effect of quantum
channels on the state of a system is a completely positive and
trace-preserving map that is described in terms of Kraus
operators.
\begin{eqnarray}
 \rho_{in}=|\psi\rangle\langle\psi|\mapsto\varepsilon_{channel}(\rho_{in})=\sum_{l} E_{l}|\psi\rangle\langle\psi|E_{l}^{\dagger}
 \end{eqnarray}
The operator $E_{l}$ satisfies the CPTP relation
$\sum_{l} E_{l}^{\dagger}E_{l} = I$.

In order to protect and improve the spin squeezing, we should perform
weak measurement $M$ and measurement reversal $N$,  before and after the decoherence channels, respectively. The two weak measurements can be written, respectively, as a non-unitary quantum operation$^{\cite{a46}}$
\begin{eqnarray}
M=\left [
 \begin {array}{cccc}
 1&0\\
 0&m
 \end {array}
 \right ]
 \qquad
 N=\left [
 \begin {array}{cccc}
 n&0\\
 0&1
 \end {array}
 \right ]
 \end{eqnarray}
where $m$ and $n$ are the measurement strengths.

After these weak measurements being implemented, the state becomes
\begin{eqnarray}
 \Theta(\rho_{in})=\frac{N\varepsilon_{channel}(M\rho_{in}M^{\dagger})N^{\dagger}}{Tr(N\varepsilon_{channel}(M\rho_{in}M^{\dagger})N^{\dagger})}
 \end{eqnarray}
 where $\varepsilon_{channel}$  is defined by Eq.(19). By discussing the symmetry of
the open system under consideration and the local decoherence and weak measurement are independent and identical. Thus, the exchange symmetry is not affected by the decoherence and weak measurement. We know that the
spin squeezing can be expressed by the local expectations
and correlations. The spin squeezing can then
calculated by the dynamics of the local expectations and
correlations. It is easy to check that an expectation value of the operator A can be calculated as
\begin{eqnarray}
 \langle A\rangle=Tr[A\Theta(\rho_{in})]=Tr[\Theta^+(A)\rho_{in}]
 \end{eqnarray}
Thus, we can calculate the expectation value via the above
equation, which is very similar to the standard Heisenberg
picture.

 \subsection{Amplitude-damping channel}
 A single qubit Kraus operators for amplitude-damping channel(ADC) is
 \begin{eqnarray}
 E_0=\sqrt{s}|0\rangle\langle0|+|1\rangle\langle1|,\;\;\;\;E_1=\sqrt{p}|1\rangle\langle0|
 \end{eqnarray}
 where $p=1-s$, $s=exp(-\gamma t/2)$ and $\gamma$ is the damping rate.

Based on the above approach and the Kraus operators
for the ADC given by Eq. (23), when $sn^2+p=m^2$, we find the evolutions of the following expectations under decoherence using weak measurement (see Appendix for details):
\begin{eqnarray}
\langle\sigma_{1z}\rangle&=&[sn^2\langle \sigma_{1z}\rangle_0-p]/M_1\\\nonumber
\langle\sigma_{1-}\sigma_{2-}\rangle&=&sm^2n^2\langle\sigma_{1-}\sigma_{2-}\rangle_0/M_1^2\\\nonumber
\langle\sigma_{1+}\sigma_{2-}\rangle&=&sm^2n^2\langle\sigma_{1+}\sigma_{2-}\rangle_0/M_1^2\\\nonumber
\langle\sigma_{1z}\sigma_{2z}\rangle&=&[s^2n^4\langle \sigma_{1z}\sigma_{2z}\rangle_0-2sn^2p\langle \sigma_{1z}\rangle_0+p^2]/M_1^2\\\nonumber
 Q_1=\langle\vec{\sigma}_1.\vec{\sigma}_2\rangle&=&[sm^2n^2+sn^2(sn^2-m^2)\langle \sigma_{1z}\sigma_{2z}\rangle_0-2sn^2p\langle \sigma_{1z}\rangle_0+p^2]/M_1^2
 \end{eqnarray}
 where $\langle\sigma_{1z}\rangle_0=-cos^{N-1}(\theta/2)$, $\langle \sigma_{1z}\sigma_{2z}\rangle_0=2^{-1}(1+cos^{N-2}(\theta))$,  $M_1=sn^2+p=m^2$.
 Substituting the relevant expectation values and the correlation function into Eqs. (9), (10), and (11) leads to
the explicit expression of the spin-squeezing parameters
\begin{eqnarray}
 \xi^2_{1}=1-sm^2n^2C_r(0)/M_1^2;
 \end{eqnarray}
 \begin{eqnarray}
 \xi^2_{2}=\frac{\xi^2_{1}}{(sn^2\langle \sigma_{1z}\rangle_0-p)/M_1)^2}
 \end{eqnarray}
  \begin{eqnarray}
  \xi^2_{3}&=&\frac{\xi^2_{1}}{(1-N^{-1})Q_1+N^{-1}}
 \end{eqnarray}
where $C_r(0)=(N-1)C_0$, $C_0=\frac{1}{4}\{[(1-cos^{N-2}\theta)^2+16sin^2(\theta/2)cos^{2N-4}(\theta/2)]^{\frac{1}{2}}-1+cos^{N-2}\theta\}$.

The expression of concurrence can be simplified to$^{\cite{a28}}$
\begin{eqnarray}
C_r&=& 2(N-1)max\{0, |u|/M_1^2-w\}
\end{eqnarray}
where $u=-\frac{1}{2}sm^2n^2Q_{12y}-sm^2n^2u_0$, $w=\frac{1}{4}(1-\langle\sigma_{1z}\sigma_{2z}\rangle)$, with $Q_{12y}=\frac{1}{2}(1-cos^{N-2}\theta)$, $u_0=-\frac{1}{8}(1-cos^{N-2}\theta)-\frac{1}{2}i\sin(\frac{\theta}{2})\cos^{N-2}(\frac{\theta}{2})$.

 In Fig. 1, we plot the spin-squeezing parameters
and concurrence against the decoherence strength $p$ under amplitude damping channel for different weak measurement strength $m$ with $\theta=0.1\pi$, $N=12$. It
clearly shows that as the decoherence strength $p$
increases, the spin squeezing decreases without weak measurement. For the smaller
value of $\theta$, there is no ESD
and SSSD. They vanish only in the asymptotic limit (see
Fig. 1(a)). However, we are able to enhance spin-squeezing parameters and the concurrence greatly by using weak measurement. Especially, they don't disappear in the asymptotic limit ( i.e. $p=1$). Moreover, with the increase of $m$, spin-squeezing parameters and the concurrence becomes a fixed value respectively. The spin-squeezing parameters and the concurrence can be completely recovered to its initial value respectively regardless of the decoherence when weak measurement strength is large (see
Fig. 1(d)). It seems that decoherence has no effect on the spin-squeezing parameters
and the concurrence. This result can be explained as
follows. According to $sn^2+p=m^2$, we have $n^2\gg1$ when the  weak measurement strength $m^2\gg1$. And, we obtain $sn^2=m^2$. From Eq.(24), we can obtain the expectations as follows
\begin{eqnarray}
\langle\sigma_{1z}\sigma_{2z}\rangle&=&\langle \sigma_{1z}\sigma_{2z}\rangle_0\\\nonumber
 Q_1&=&\langle\vec{\sigma}_1.\vec{\sigma}_2\rangle=1
 \end{eqnarray}
 Thus, the spin-squeezing parameters and concurrence can be calculated as
\begin{eqnarray}
\xi^2_{1}&=&1-C_r(0)\\\nonumber
 \xi^2_{2}&=&\frac{\xi^2_{1}}{\langle \sigma_{1z}\rangle_0^2}\\\nonumber
 \xi^2_{3}&=&\xi^2_{1}\\\nonumber
 C_r&=&\zeta^2_3=C_r(0)
\end{eqnarray}
So, the spin-squeezing parameters and the concurrence can be completely recovered to its initial value when weak measurement strength is very large. The overlap of the solid line and the dashed line in Fig. 1(d) due to the fact that the spin squeezing $\zeta^2_3$ and the concurrence $C_r(0)$ are equivalent for the initial state Eq.(2).

We plot the spin-squeezing parameters
and concurrence against the decoherence strength $p$ under amplitude damping channels for different weak measurement strength $m$ with $\theta=1.8\pi$, $N=12$ in Fig. 2. For larger values of $\theta$, as the decoherence strength $p$
increases, the spin squeezing decreases until it suddenly
vanishes, so the phenomenon of ESD and SSSD occurs when there is no weak measurement (see Fig. 2(a)).
However, the spin-squeezing parameters and concurrence can be improved greatly by using weak measurement.  Moreover, with the increase of $m$, the phenomenon of ESD and SSSD can be avoided. When the measurement strength $m$ is very large, the spin-squeezing parameters and  the concurrence can be completely recovered to its initial value respectively no matter what the decoherence parameter is (see Fig. 2(d)).

 \subsection{Depolarizing channel}
  A single qubit Kraus operators for depolarizing channel(DPC) is
  \begin{eqnarray}
 E_0=\sqrt{1-p'}I,\;\;\;\;\;E_1=\sqrt{\frac{p'}{3}}\sigma_x\\\nonumber
 E_2=\sqrt{\frac{p'}{3}}\sigma_y,\;\;\;\;\;E_3=\sqrt{\frac{p'}{3}}\sigma_z
 \end{eqnarray}
 where $p'=3p/4$ and $I$ is the identity operator.

From Eq.(22) and the Kraus operators
for the DPC given by Eq. (31),  when $m=1$, we find the evolutions of the following expectations under decoherence using weak measurement (see Appendix for details):
\begin{eqnarray}
\langle\sigma_{1z}\rangle&=&\frac{1}{2}[(n^2s+s)\langle \sigma_{1z}\rangle_0+(n^2-1)]/M_2\\\nonumber
\langle\sigma_{1-}\sigma_{2-}\rangle&=&s^2n^2\langle\sigma_{1-}\sigma_{2-}\rangle_0/M_2^2\\\nonumber
\langle\sigma_{1+}\sigma_{2-}\rangle&=&s^2n^2\langle\sigma_{1+}\sigma_{2-}\rangle_0/M_2^2\\\nonumber
\langle\sigma_{1z}\sigma_{2z}\rangle&=&\frac{1}{4}[(n^2s+s)^2\langle \sigma_{1z}\sigma_{2z}\rangle_0+2(n^2-1)(n^2s+s)\langle \sigma_{1z}\rangle_0+(n^2-1)^2]/M_2^2\\\nonumber
Q_2&=&\langle\vec{\sigma}_1.\vec{\sigma}_2\rangle=\{s^2n^2(1-\langle \sigma_{1z}\sigma_{2z}\rangle_0)+\frac{1}{4}[(n^2s+s)^2\langle \sigma_{1z}\sigma_{2z}\rangle_0\\\nonumber
  &+&2(n^2-1)(n^2s+s)\langle \sigma_{1z}\rangle_0+(n^2-1)^2]\}/M_2^2
 \end{eqnarray}
 where $M_2=\frac{1}{2}[(n^2s-s)\langle \sigma_{1z}\rangle_0+(n^2+1)]$.
 Substituting the relevant expectation values and the correlation function into Eqs. (9), (10), and (11) leads to
the explicit expression of the spin-squeezing parameters
\begin{eqnarray}
 \xi^2_{1}=1-s^2n^2C_r(0)/M_2^2;
 \end{eqnarray}
 \begin{eqnarray}
 \xi^2_{2}= \frac{\xi^2_{1}}{\{\frac{1}{2}[(n^2s+s)\langle \sigma_{1z}\rangle_0+(n^2-1)]/M_2\}^2}
 \end{eqnarray}
  \begin{eqnarray}
  \xi^2_{3}&=&\frac{\xi^2_{1}}{(1-N^{-1})Q_2+N^{-1}}
 \end{eqnarray}

The expression of concurrence can be simplified to$^{\cite{a28}}$
\begin{eqnarray}
C_r&=& 2(N-1)max\{0, |u|/M_2^2-w\}
\end{eqnarray}
where, $u=-\frac{1}{2}s^2n^2Q_{12y}-s^2n^2u_0$.

 In Fig.3, we plot the spin-squeezing parameters
and concurrence against the decoherence strength $p$ under depolarizing channel with $\theta=1.8\pi$, $N=12$. We can see that similar to amplitude damping channel, the spin squeezing decreases as the decoherence strength p
increases without weak measurement. And, the phenomenon of ESD and SSSD occurs (see Fig.3(a)). However, we are able to improve the spin-squeezing parameters $\zeta^2_{3}$ and the concurrence greatly by using weak measurement. The larger is the weak measurement strength $n$, the later is the vanishing time. And when weak measurement strength is very large, the spin-squeezing parameter $\zeta^2_{3}$ and the concurrence vanish approximately in the asymptotic limit (see Fig.3(d)). We note that with the increase of weak measurement strength $n$, the spin-squeezing parameter $\zeta^2_{2}$ becomes more and more weak until it is zero. This means that in our model, the parameter $\xi^2_{3}<1$ implies the existence of
pairwise entanglement, while $\xi^2_{2}<1$ does not.

\subsection{Phase-damping channel}
A single qubit Kraus operators for phase-damping channel (PDC) is
\begin{eqnarray}
 E_0=\sqrt{s}I,\;\;\;E_1=\sqrt{p}|0\rangle\langle0|, \;\;\;E_2=\sqrt{p}|1\rangle\langle1|
 \end{eqnarray}

From Eq.(22) and the Kraus operators
for the PDC given by Eq. (37), when $n^2-1=m^2+1$, we find the evolutions of the following expectations under decoherence using weak measurement (see Appendix for details):
\begin{eqnarray}
\langle\sigma_{1z}\rangle&=&[(m^2+1)\langle \sigma_{1z}\rangle_0+1]/M_3\\\nonumber
\langle\sigma_{1-}\sigma_{2-}\rangle&=&s^2m^2n^2\langle\sigma_{1-}\sigma_{2-}\rangle_0/M_3^2\\\nonumber
\langle\sigma_{1+}\sigma_{2-}\rangle&=&s^2m^2n^2\langle\sigma_{1+}\sigma_{2-}\rangle_0/M_3^2\\\nonumber
\langle\sigma_{1z}\sigma_{2z}\rangle&=&[(m^2+1)^2\langle \sigma_{1z}\sigma_{2z}\rangle_0+2(m^2+1)\langle \sigma_{1z}\rangle_0+1]/M_3^2\\\nonumber
 Q_3=\langle\vec{\sigma}_1.\vec{\sigma}_2\rangle &=&[s^2m^2n^2(1-\langle \sigma_{1z}\sigma_{2z}\rangle_0)+(m^2+1)^2\langle \sigma_{1z}\sigma_{2z}\rangle_0+2(m^2+1)\langle \sigma_{1z}\rangle_0+1]/M_3^2\\\nonumber
 \end{eqnarray}
 where $M_3=m^2+1+\langle\sigma_{1z}\rangle_0$.
 Substituting the relevant expectation values and the correlation function into Eqs. (9), (10), and (11) leads to
the explicit expression of the spin-squeezing parameters
\begin{eqnarray}
 \xi^2_{1}=1-s^2m^2n^2C_r(0)/M_3^2;
 \end{eqnarray}
 \begin{eqnarray}
 \xi^2_{2}=\frac{\xi^2_{1}}{((m^2+1)\langle\sigma_{1z}\rangle_0+1)/M_3)^2}
 \end{eqnarray}
  \begin{eqnarray}
  \xi^2_{3}&=&\frac{\xi^2_{1}}{(1-N^{-1})Q_3+N^{-1}}
 \end{eqnarray}

The expression of concurrence can be simplified to$^{\cite{a28}}$
\begin{eqnarray}
C_r&=& 2(N-1)max\{0, |u|/M_3^2-w\}
\end{eqnarray}
where, $u=-\frac{1}{2}s^2m^2n^2Q_{12y}-s^2m^2n^2u_0$.

 In Fig.4, we plot the spin-squeezing parameters
and concurrence against the decoherence strength $p$ under phase-damping channel with $\theta=1.8\pi$, $N=12$. We can see that similar to amplitude- damping and depolarizing channel, the spin squeezing decreases as the decoherence strength $p$
increases without weak measurement. And the phenomenon of ESD and SSSD occurs (see Fig.4(a)). However, we are able to enhance the spin-squeezing parameters $\zeta^2_{3}$ and the concurrence greatly, and to avoid the phenomenon of ESD and SSSD by using weak measurement. Morover, when weak measurement strength $m$ is small, the spin-squeezing parameter $\zeta^2_{3}$ and the concurrence becomes a fixed value respectively regardless of the decoherence  although the spin-squeezing parameter $\zeta^2_{2}$ becomes zero(see Fig.4(d)). This result can be explained as
follows. When the  weak measurement strength $m^2\ll1$, according to $n^2-1=m^2+1$, we have $n^2=2$. So, we obtain $M_3=1+\langle\sigma_{1z}\rangle_0$ and $s^2m^2n^2\ll M_3^2$. From Eq.(38), we can obtain the expectations as follows
\begin{eqnarray}
\langle\sigma_{1z}\sigma_{2z}\rangle&=&[\langle \sigma_{1z}\sigma_{2z}\rangle_0+2\langle \sigma_{1z}\rangle_0+1]/M_3^2\\\nonumber
 Q_3=\langle\vec{\sigma}_1.\vec{\sigma}_2\rangle &=&[\langle \sigma_{1z}\sigma_{2z}\rangle_0+2\langle \sigma_{1z}\rangle_0+1]/M_3^2\\\nonumber
 \end{eqnarray}
 Thus, the spin-squeezing parameters and concurrence can be calculated as
\begin{eqnarray}
\xi^2_{1}&=&1\\\nonumber
 \xi^2_{2}&=&\xi^2_{1}=1\\\nonumber
 \xi^2_{3}&=&\frac{1}{(1-N^{-1})Q_3+N^{-1}}\\\nonumber
 C_r&=&\frac{1}{2}(N-1)\{[\langle \sigma_{1z}\sigma_{2z}\rangle_0+2\langle \sigma_{1z}\rangle_0+1]/M_3^2-1\}
\end{eqnarray}
So, the spin-squeezing parameter $\zeta^2_{3}$ and the concurrence can be recovered to certain stationary value respectively and the spin-squeezing parameter $\zeta^2_{2}=0$ when weak measurement strength $m$ is very small.

We also note that with the decrease of weak measurement strength $m$, the spin-squeezing parameter $\zeta^2_{2}$ becomes more and more weak until it is zero. This means that in our model, the parameter $\xi^2_{3}<1$ implies the existence of
pairwise entanglement, while $\xi^2_{2}<1$ does not. This result is the same as that discussed in the case of depolarizing channel.

\section{Conclusion}
In this paper, we have proposed an efficient method to protect spin squeezing under the action of amplitude-damping, depolarizing and phase-damping channels  based on measurement reversal from weak measurement, and have considered an ensemble of N independent
spin-1/2 particles with exchange symmetry. We have found that spin squeezing can be enhanced greatly under three different decoherence channels and spin-squeezing sudden death can be avoided undergoing amplitude-damping and phase-damping channels.

{\bf Acknowledgement}

This work is supported by the National Natural Science Foundation of
China (Grant No.11374096) and
the Natural Science Foundation of Hunan Province
 of China (Grant No. 2016JJ2014).

\newpage
{\bf Appendix: Derivation of the evolution of the
correlations and expectations under decoherence by using weak measurements}

For an arbitrary matrix
\begin{eqnarray}
A=\left [
 \begin {array}{cccc}
 a&b\\
 c&d
 \end {array}
 \right ],
 \end{eqnarray}
from Eq.(22) and the Kraus operators (23) for the ADC, when $sn^2+p=m^2$, it is straight forward to find
\begin{eqnarray}
\Theta^{+}(A)=\left [
 \begin {array}{cccc}
 asn^2+dp&bmn\sqrt{s} \\
 cmn\sqrt{s}&dm^2
 \end {array}
 \right ]/(sn^2+p),
\end{eqnarray}
The above equation imply that
\begin{eqnarray}
\Theta^+(\sigma_\mu)=mn\sqrt{s}\sigma_\mu /(sn^2+p) \;\;\;\;for\;\; \mu=x,y\\
\Theta^+(\sigma_z)=(sn^2\sigma_z-p)/(sn^2+p)
\end{eqnarray}
As we considered independent and identical decoherence
channels and weak measurements acting separately on each spin, the evolution
correlations and expectations in Eq. (24), are obtained directly from the above equations.

From Eqs.(31) and (22), when $m=1$, the evolution of the
matrix A under the DPC is obtained as
\begin{eqnarray}
\Theta^{+}(A)=\left [
 \begin {array}{cccc}
 \frac{d}{2}p+an^2-\frac{a}{2}n^2p&bns \\
 cns&\frac{ap}{2}n^2+d-\frac{d}{2}p
 \end {array}
 \right ]/[\frac{1}{2}(n^2+1)+\frac{1}{2}(n^2s-s)\langle\sigma_z\rangle_0],
\end{eqnarray}
from which one finds
 \begin{eqnarray}
\Theta^+(\sigma_\mu)&=&ns \sigma_\mu /[\frac{1}{2}(n^2+1)+\frac{1}{2}(n^2s-s)\langle\sigma_z\rangle_0] \;\;\;\;for\;\; \mu=x,y\\
\Theta^+(\sigma_z)&=&[\frac{1}{2}(n^2s+s)\sigma_z+\frac{1}{2}(n^2-1)]/[\frac{1}{2}(n^2+1)+\frac{1}{2}(n^2s-s)\langle\sigma_z\rangle_0]
\end{eqnarray}

From Eqs.(37) and (22), when $n^2-1=m^2+1$, the evolution of the
matrix A under the PDC is obtained as
\begin{eqnarray}
\Theta^{+}(A)=\left [
 \begin {array}{cccc}
 an^2&bmns \\
 cmns&dm^2
 \end {array}
 \right ]/[(m^2+1)+\langle\sigma_z\rangle_0],
\end{eqnarray}
from which one finds
 \begin{eqnarray}
\Theta^+(\sigma_\mu)&=&mns \sigma_\mu /[(m^2+1)+\langle\sigma_z\rangle_0] \;\;\;\;for\;\; \mu=x,y\\
\Theta^+(\sigma_z)&=&[(m^2+1)\sigma_z+1]/[(m^2+1)+\langle\sigma_z\rangle_0]
\end{eqnarray}

\begin{figure}[htb]
 \centering
\includegraphics[width=2.7in]{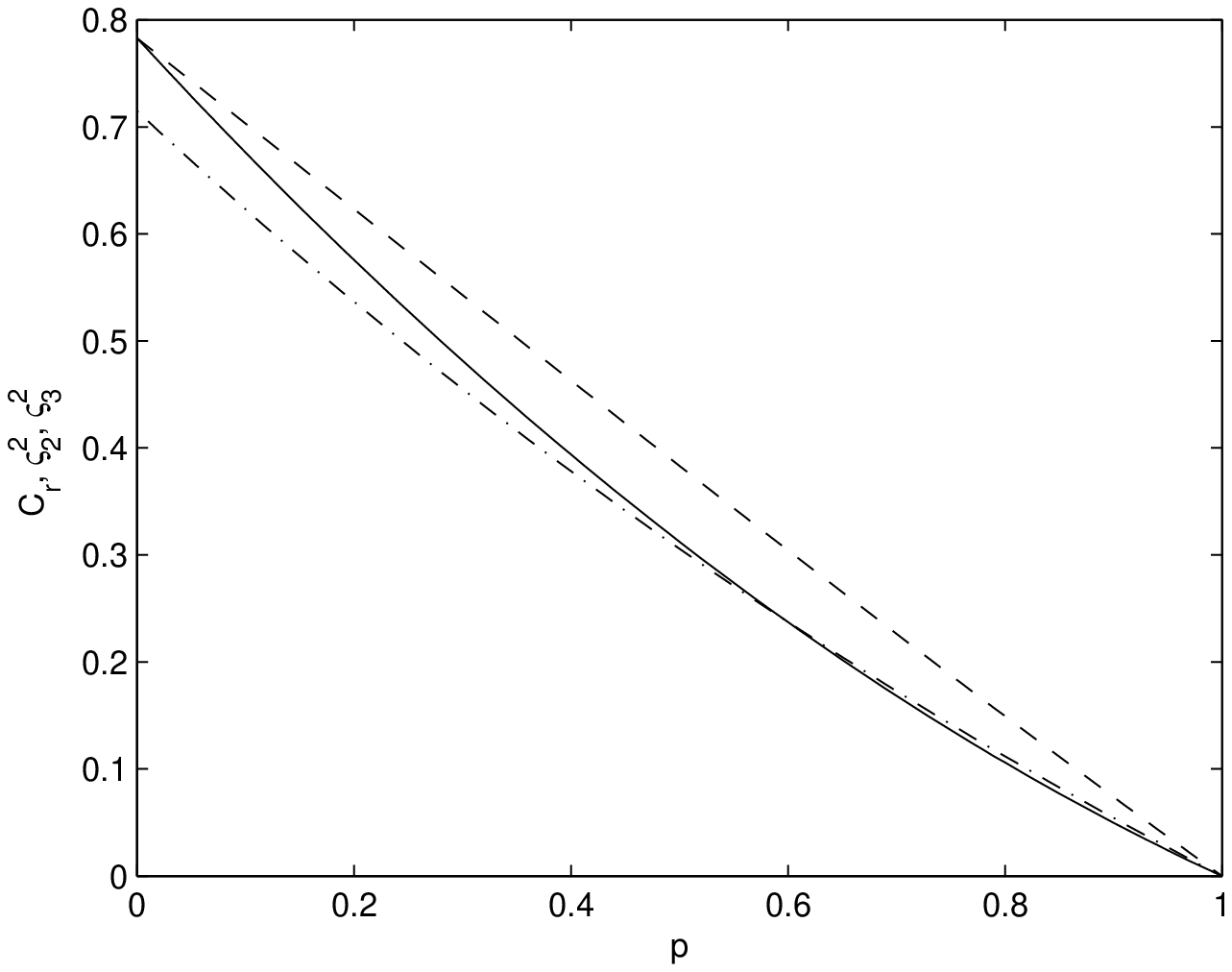}(a)
\hspace{0.01mm}
\includegraphics[width=2.7in]{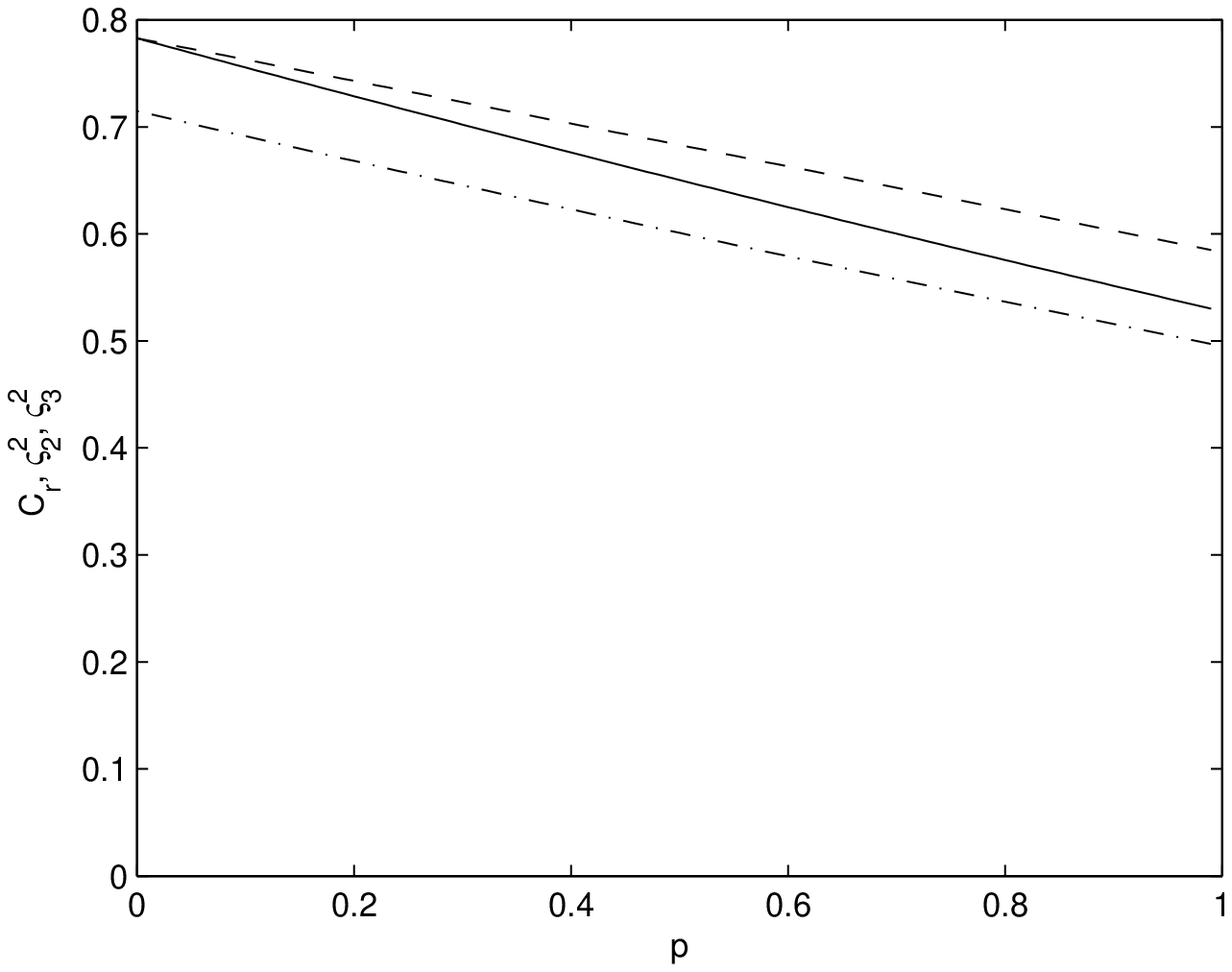}(b)
\hspace{0.01mm}
\includegraphics[width=2.7in]{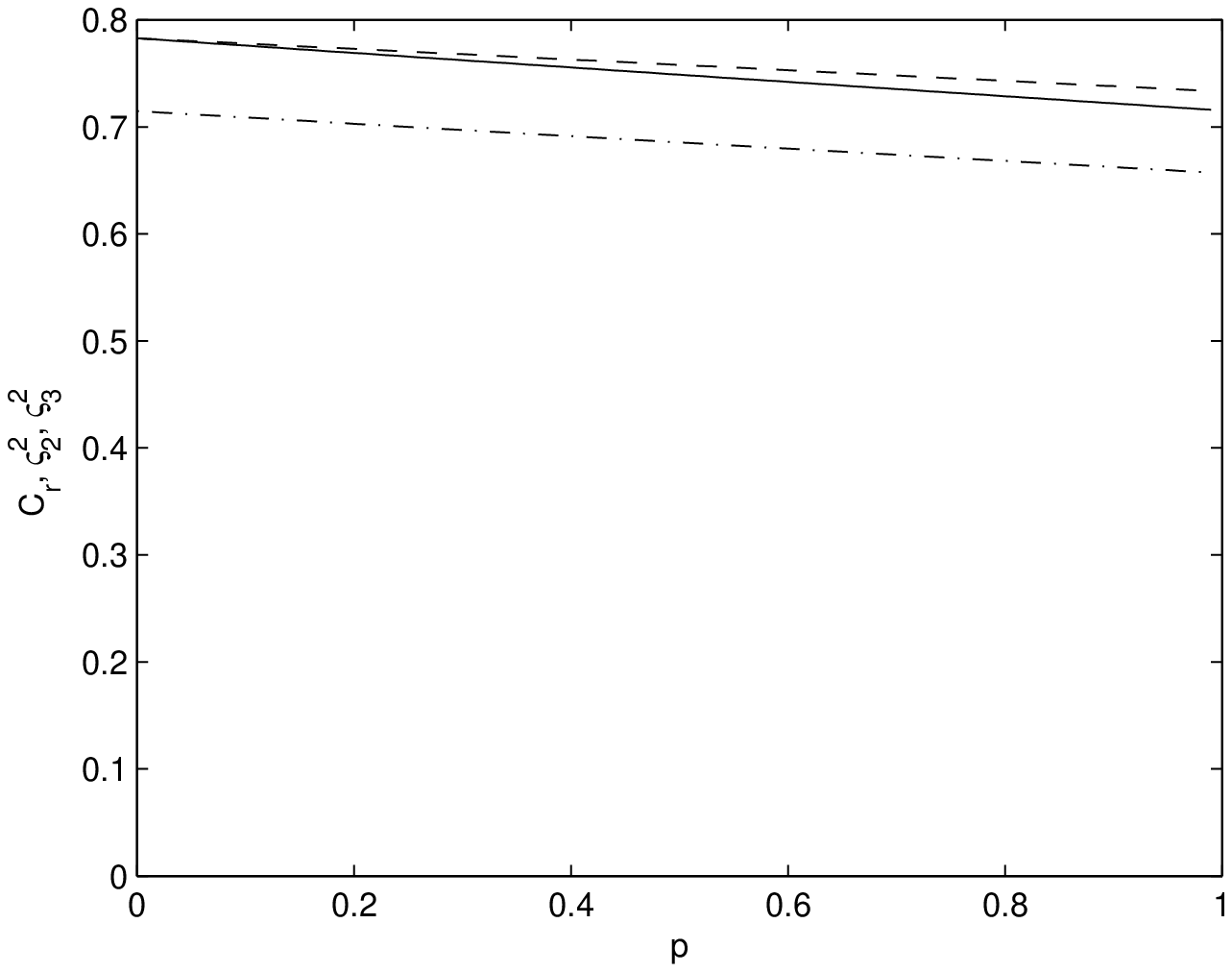}(c)
\hspace{0.01mm}
\includegraphics[width=2.7in]{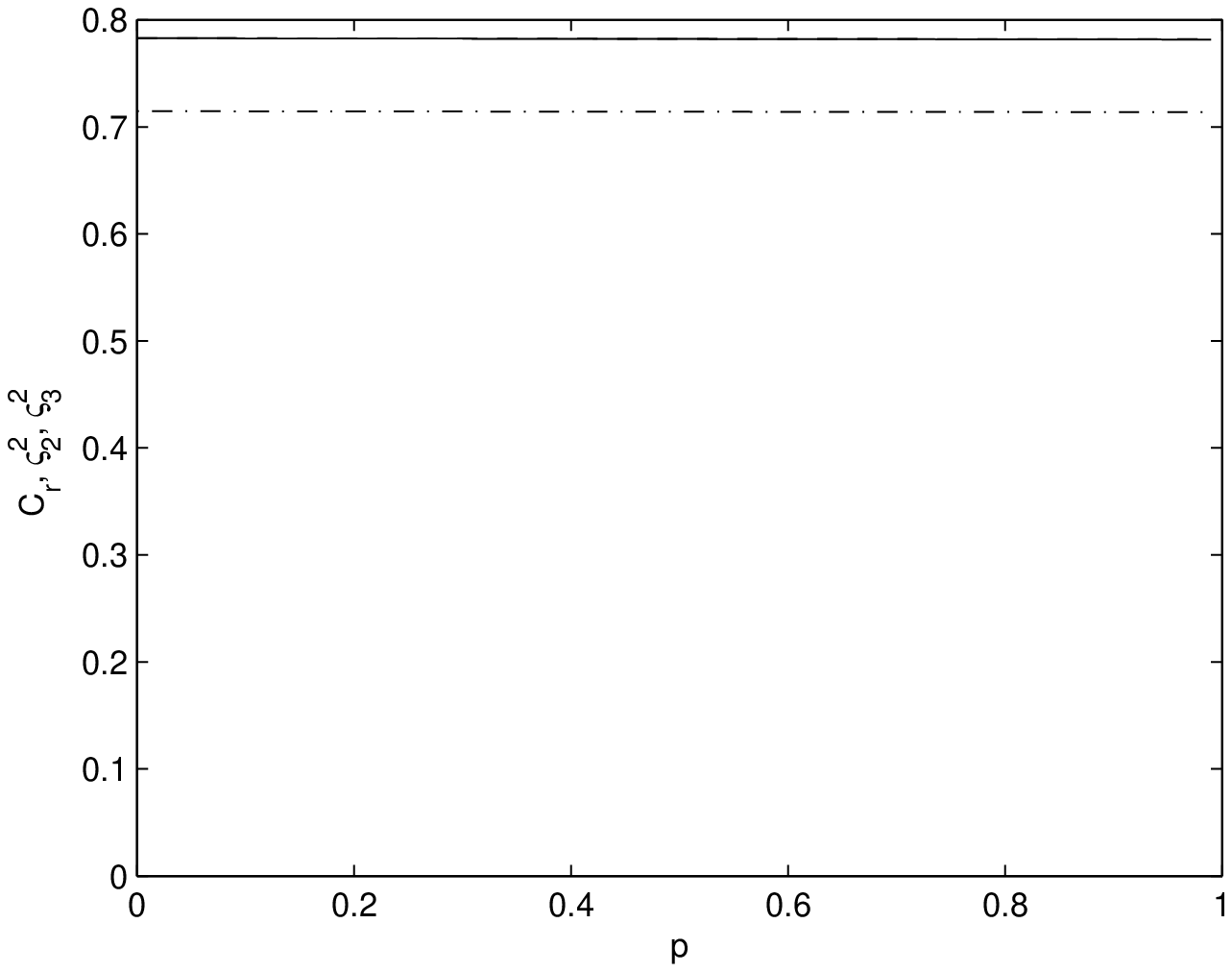}(d)
\hspace{0.01mm}
\caption{ Spin-squeezing parameters $\varsigma_2^2$ (dash-dotted line), $\varsigma_3^2$ (dashed line)
and the concurrence $C_r$ (solid line) versus the decoherence strength $p$ for the amplitude-damping channel with $\theta=0.1\pi$, $N=12$. (a) Without weak measurement; (b) weak measurement strength $m=2$; (c) $m=4$; (d) $m=30$.}
  \end{figure}

 \begin{figure}[htb]
 \centering
\includegraphics[width=2.7in]{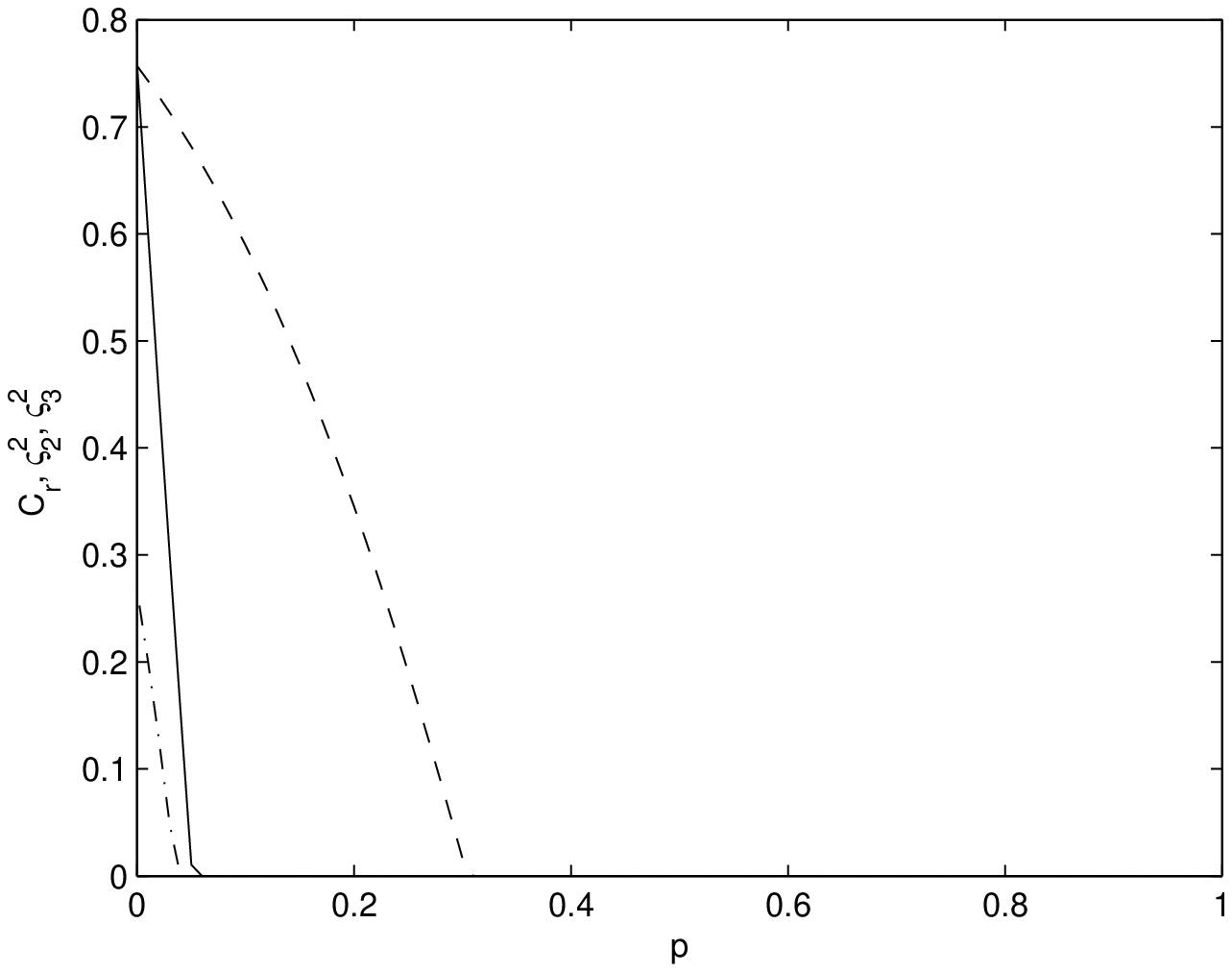}(a)
\hspace{0.01mm}
\includegraphics[width=2.7in]{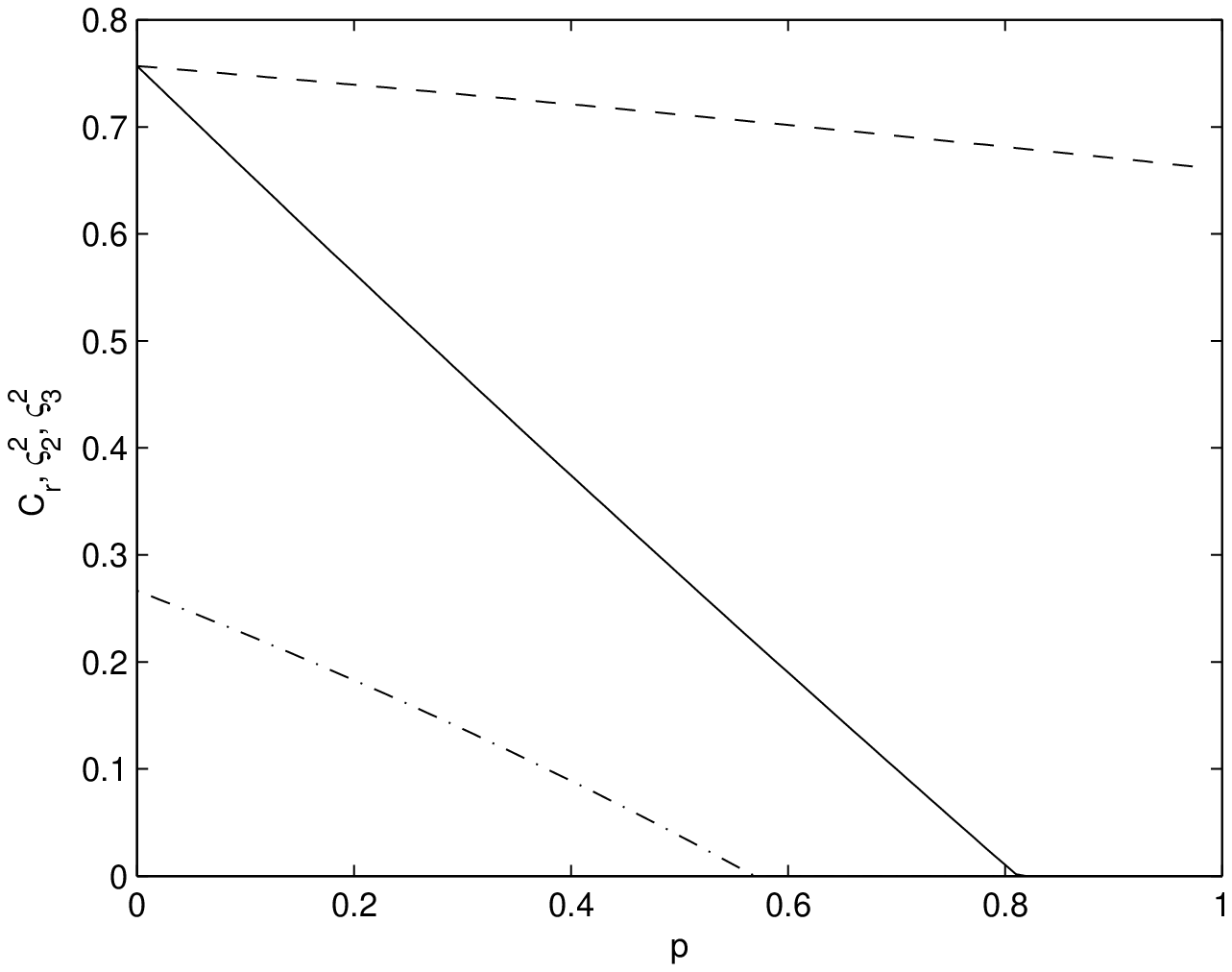}(b)
\hspace{0.01mm}
\includegraphics[width=2.7in]{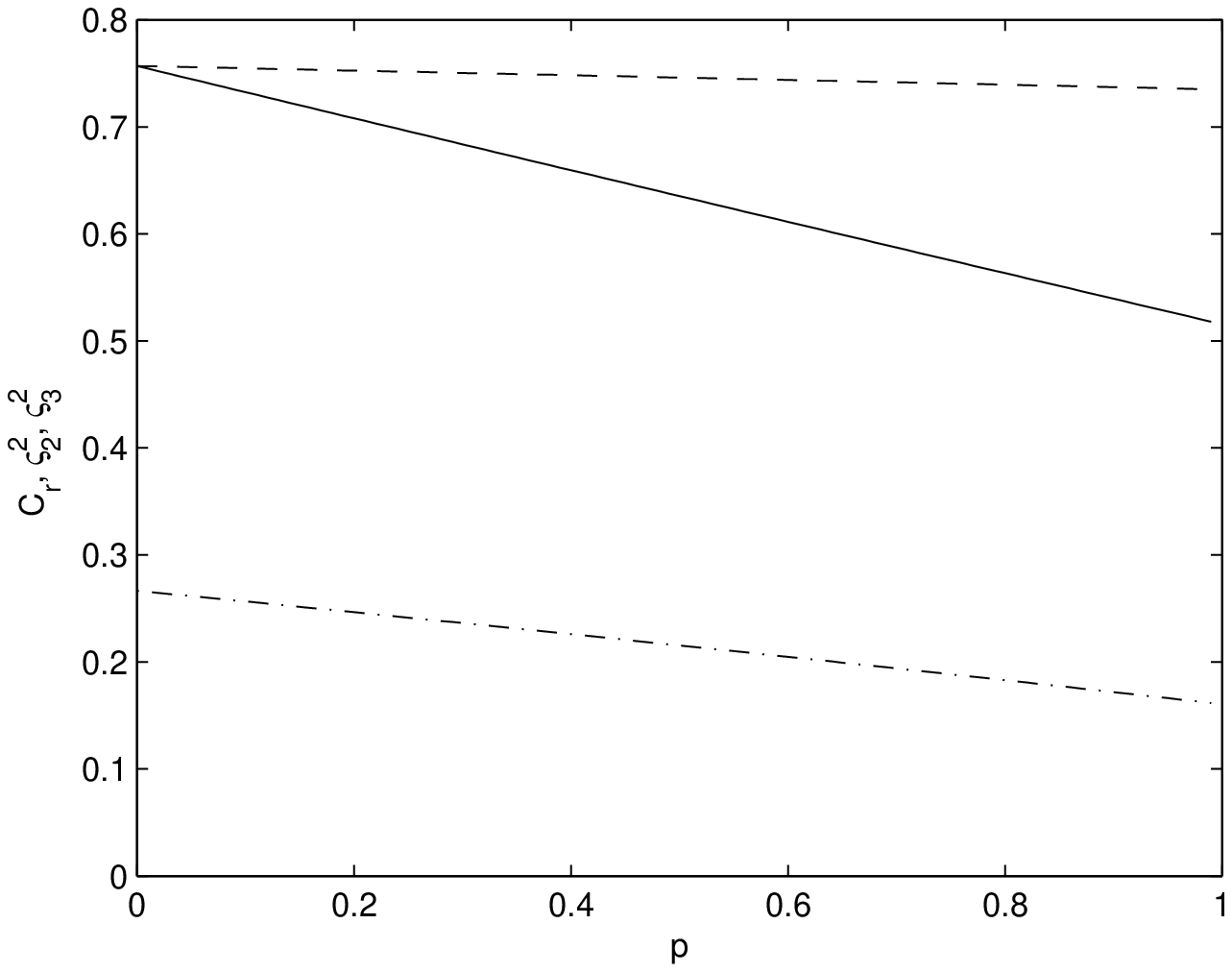}(c)
\hspace{0.01mm}
\includegraphics[width=2.7in]{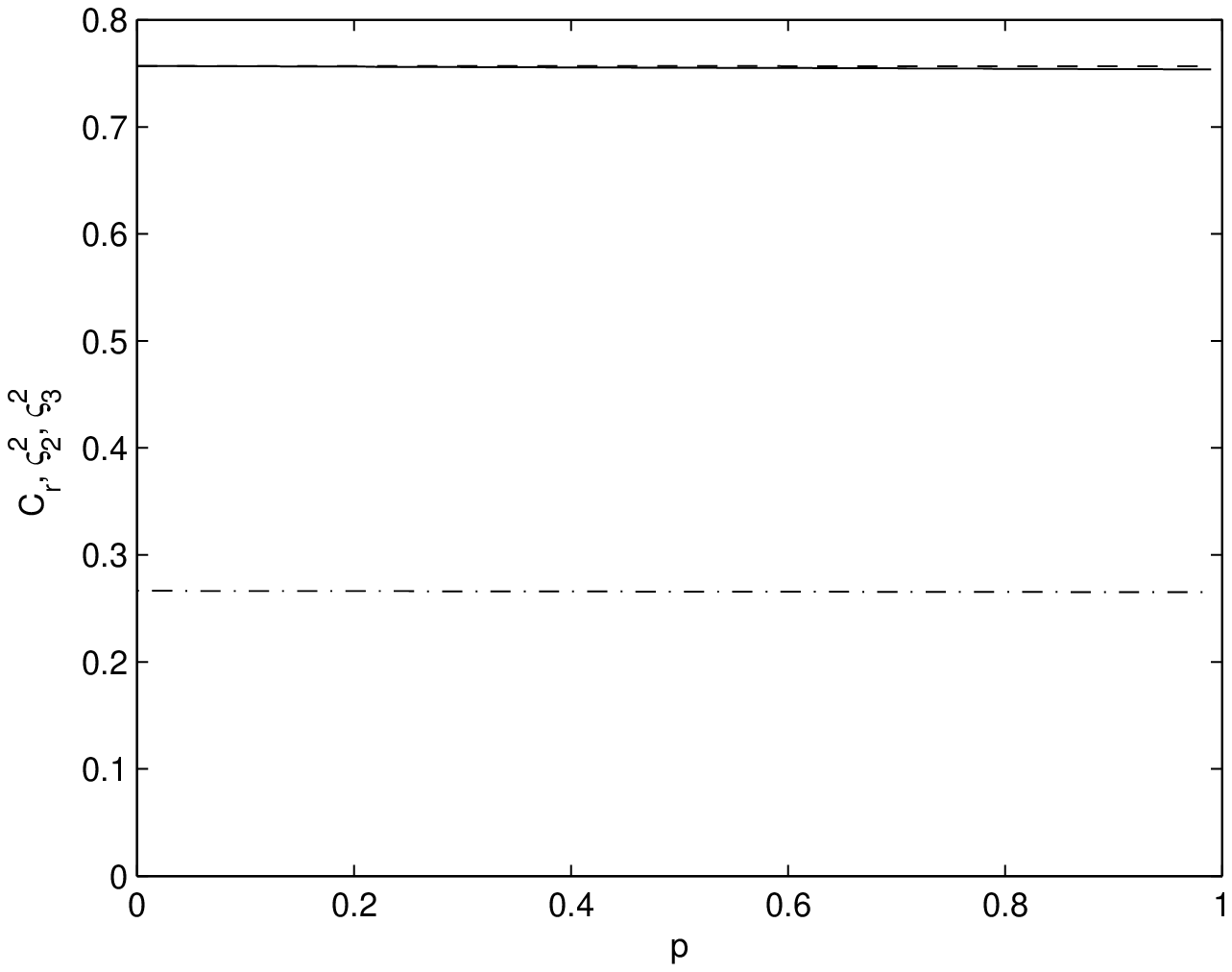}(d)
\hspace{0.01mm}
\caption{ Spin-squeezing parameters $\varsigma_2^2$ (dash-dotted line), $\varsigma_3^2$ (dashed line)
and the concurrence $C_r$ (solid line) versus the decoherence strength $p$ for the amplitude-damping channel with $\theta=1.8\pi$, $N=12$. (a) Without weak measurement; (b) weak measurement strength $m=4$; (c) $m=8$; (d) $m=70$.}
  \end{figure}

\begin{figure}[htb]
 \centering
\includegraphics[width=2.7in]{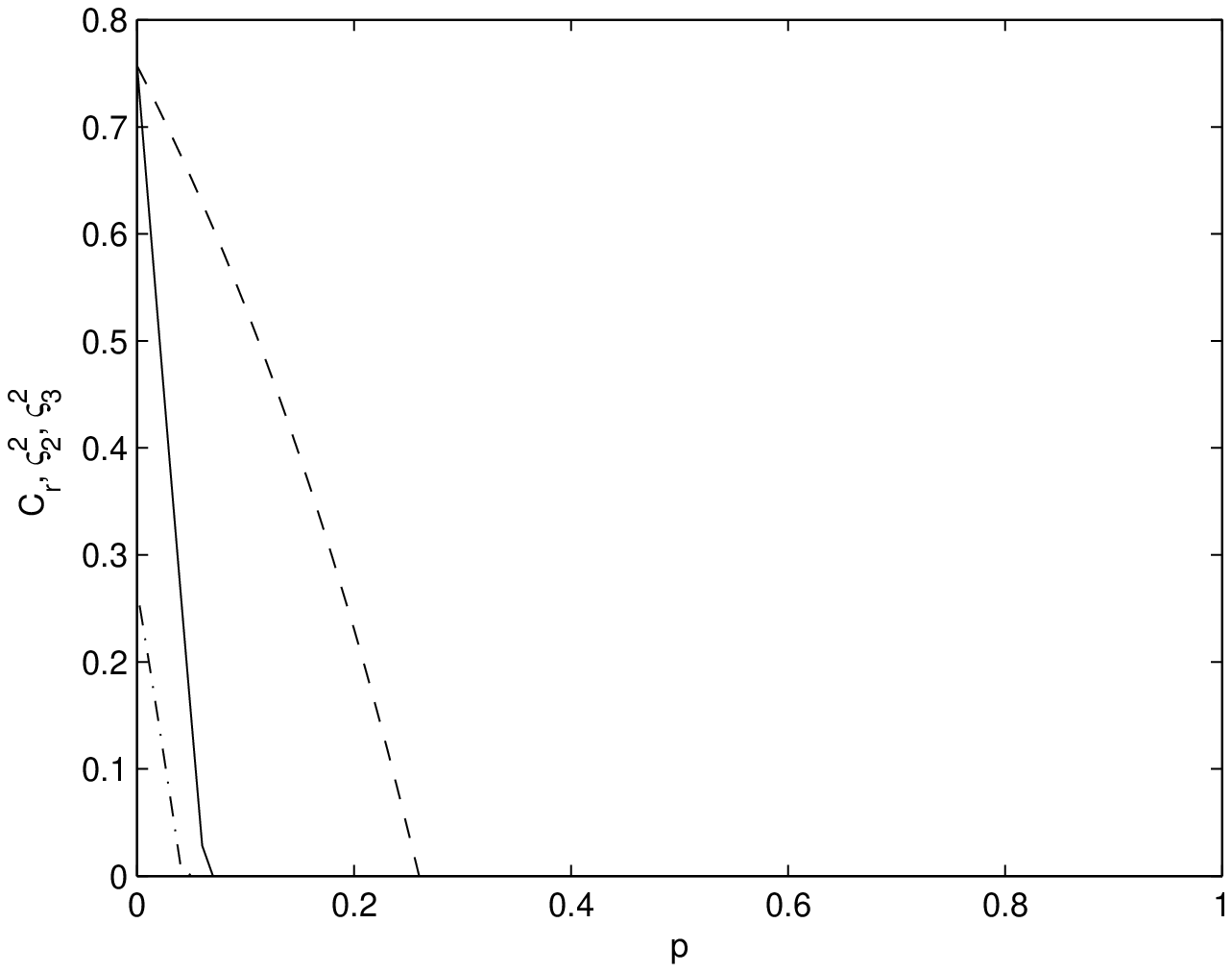}(a)
\hspace{0.01mm}
\includegraphics[width=2.7in]{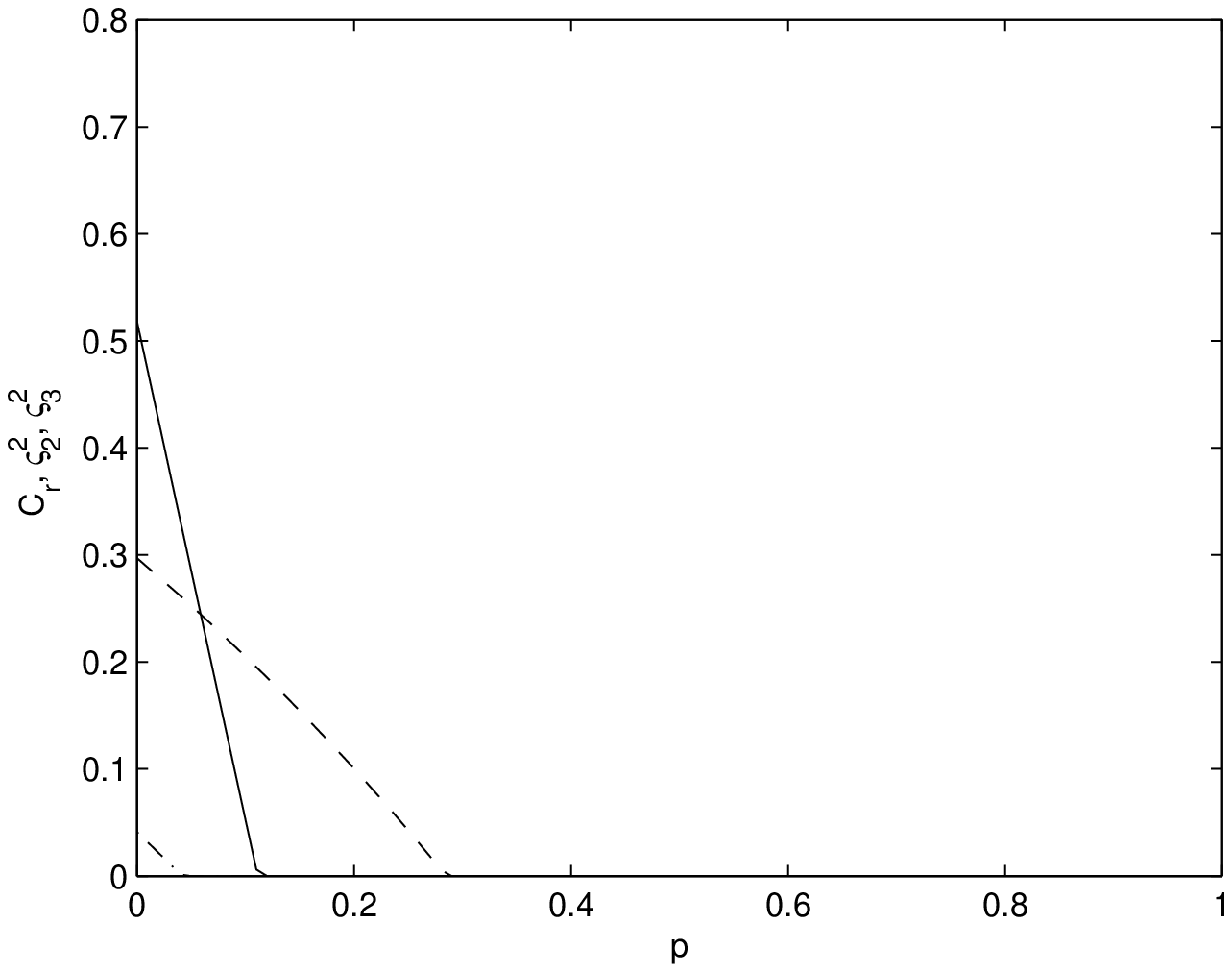}(b)
\includegraphics[width=2.7in]{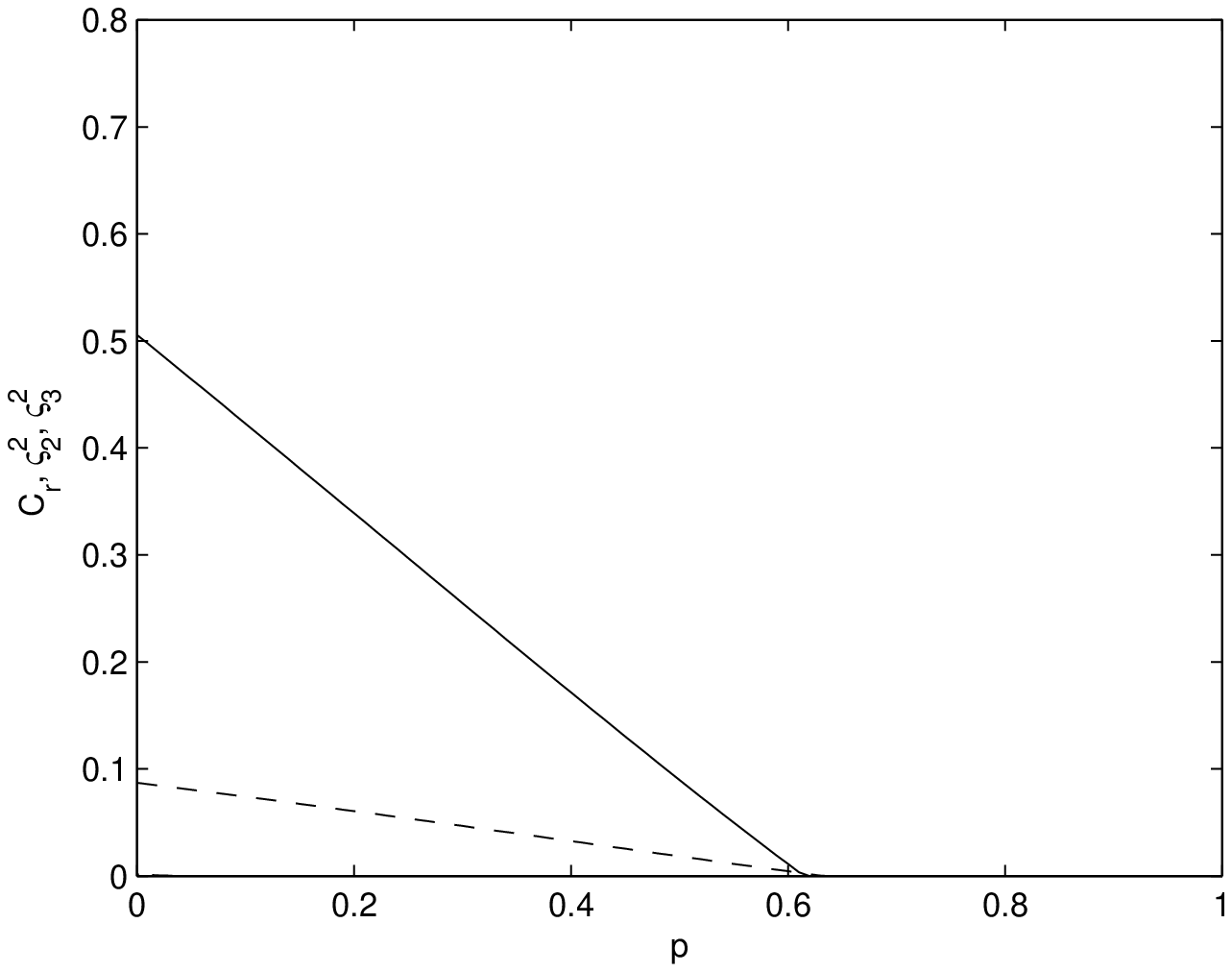}(c)
\hspace{0.01mm}
\includegraphics[width=2.7in]{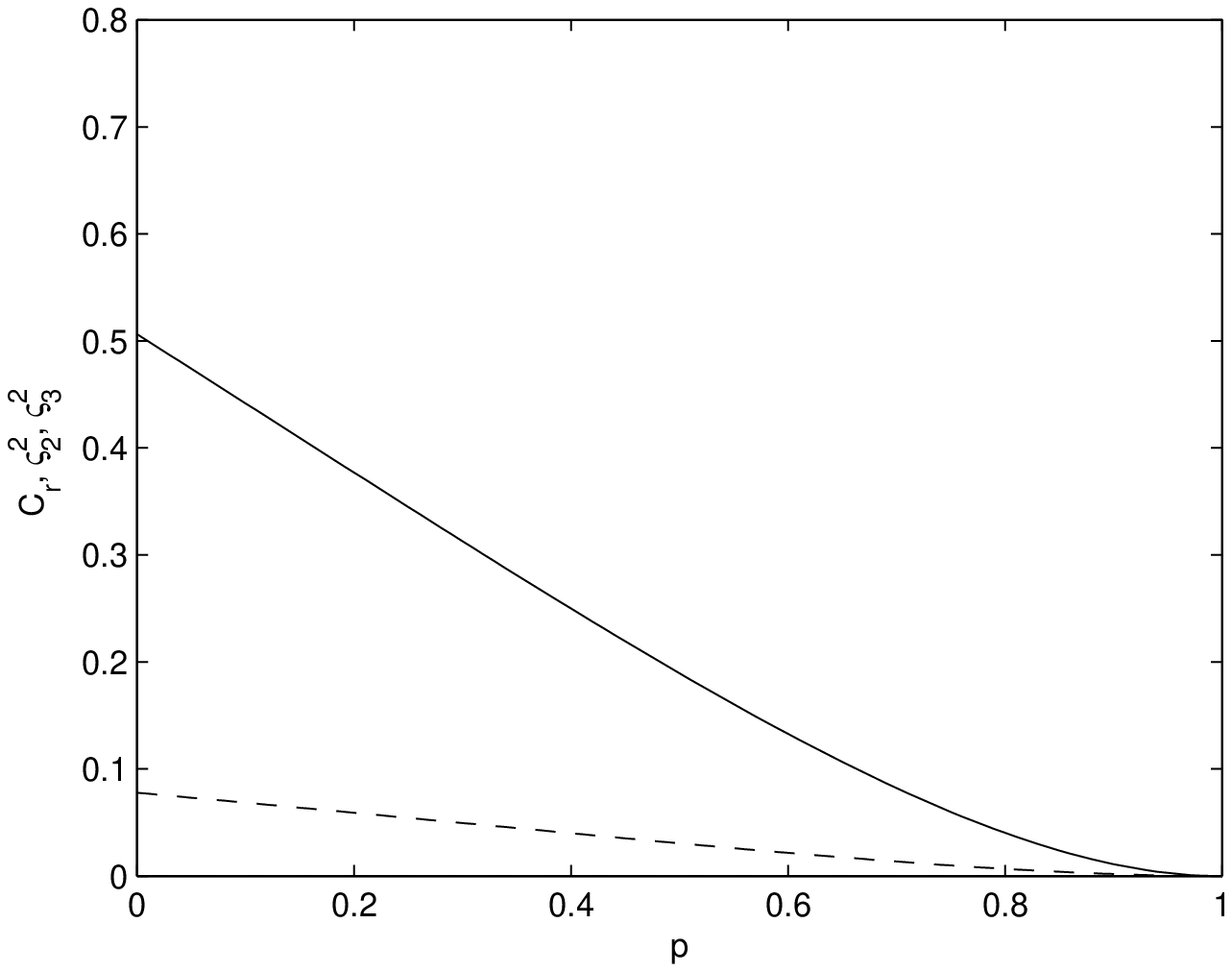}(d)
\hspace{0.01mm}
\caption{ Spin-squeezing parameters $\varsigma_2^2$ (dash-dotted line), $\varsigma_3^2$ (dashed line)
and the concurrence $C_r$ (solid line) versus the decoherence strength $p$ for the depolarizing channel with $\theta=1.8\pi$, $N=12$. (a) Without weak measurement; (b) weak measurement strength  $n=2$; (c) $n=10$; (d) $n=500$.}
  \end{figure}

\begin{figure}[htb]
 \centering
\includegraphics[width=2.7in]{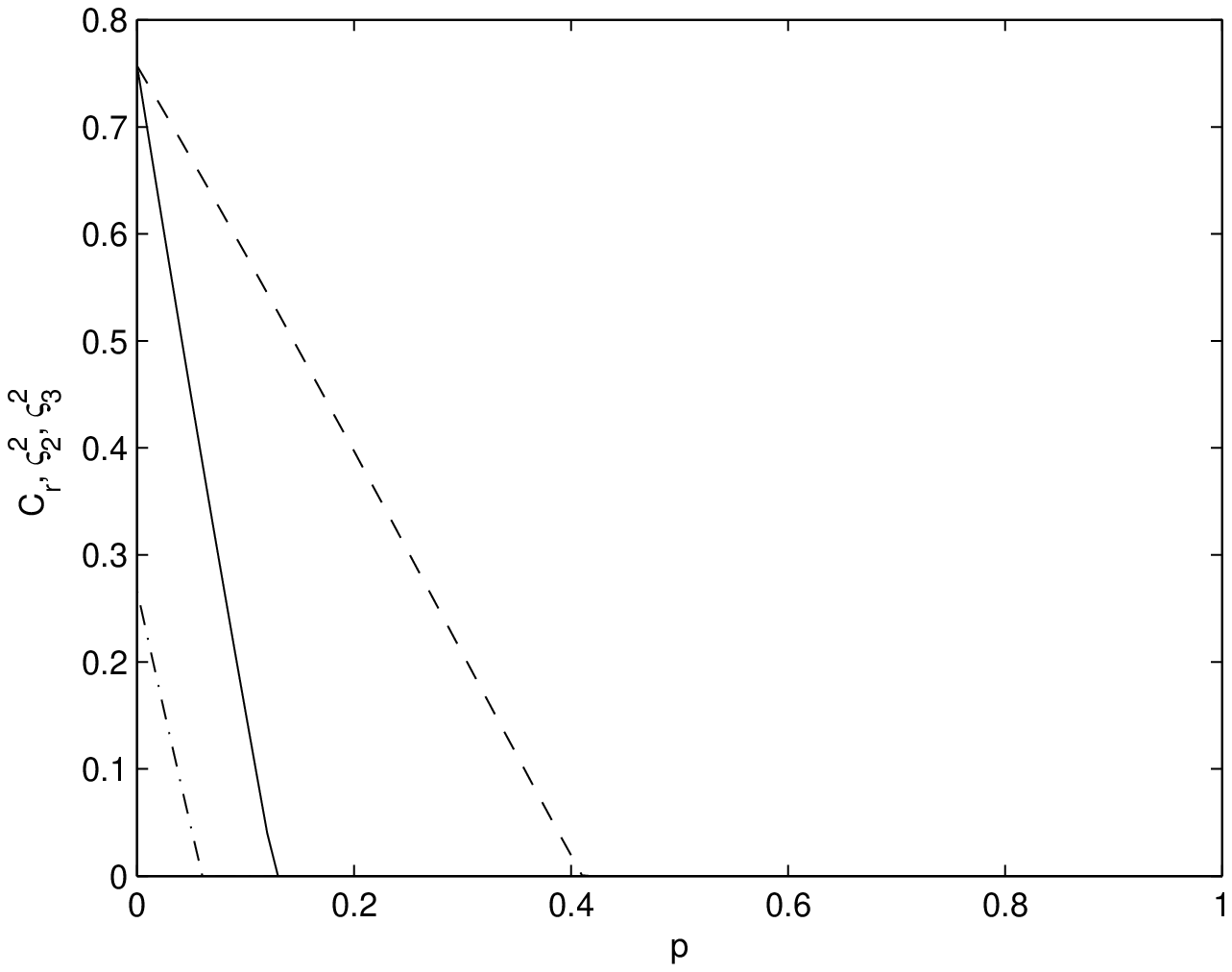}(a)
\hspace{0.01mm}
\includegraphics[width=2.7in]{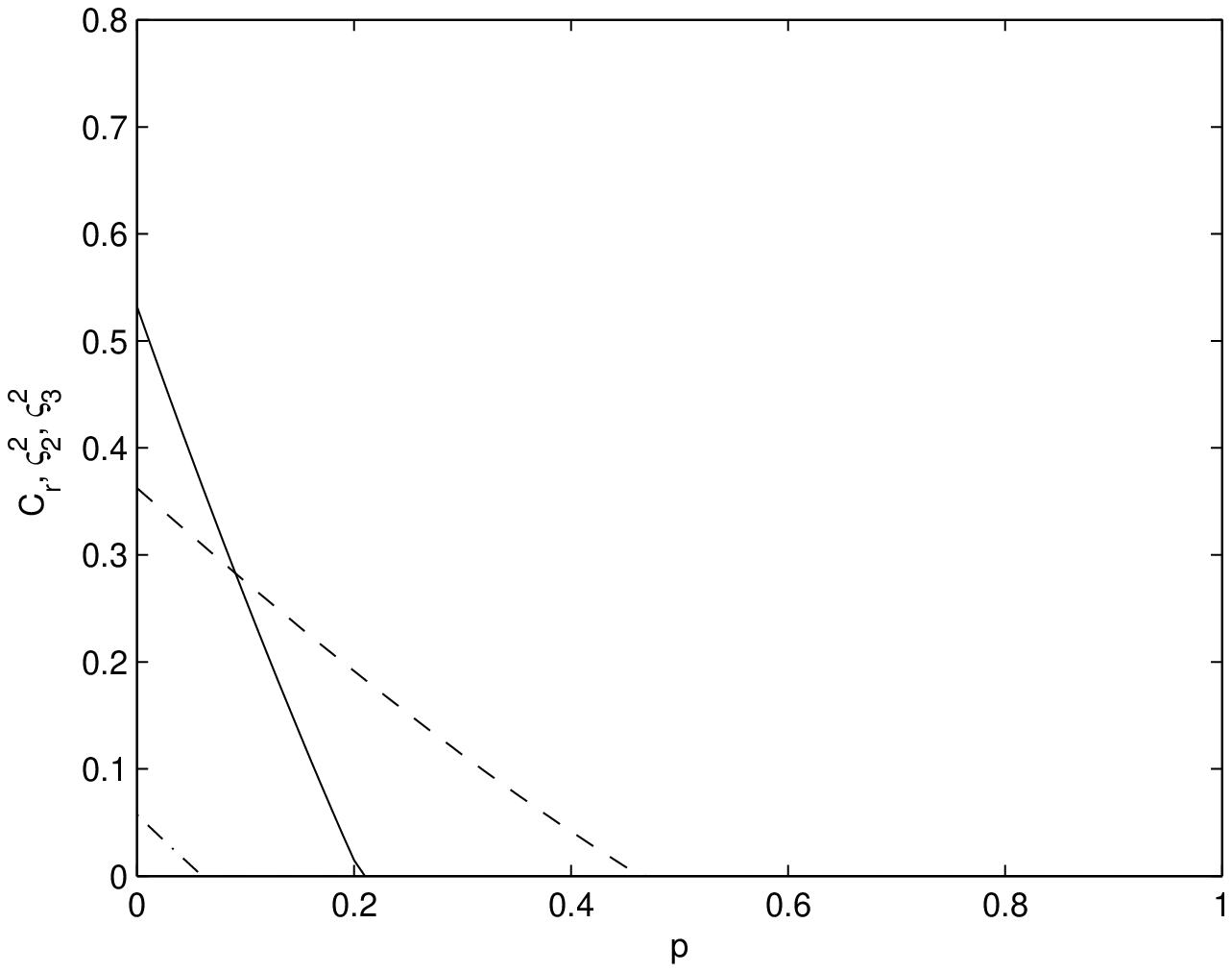}(b)
\hspace{0.01mm}
\includegraphics[width=2.7in]{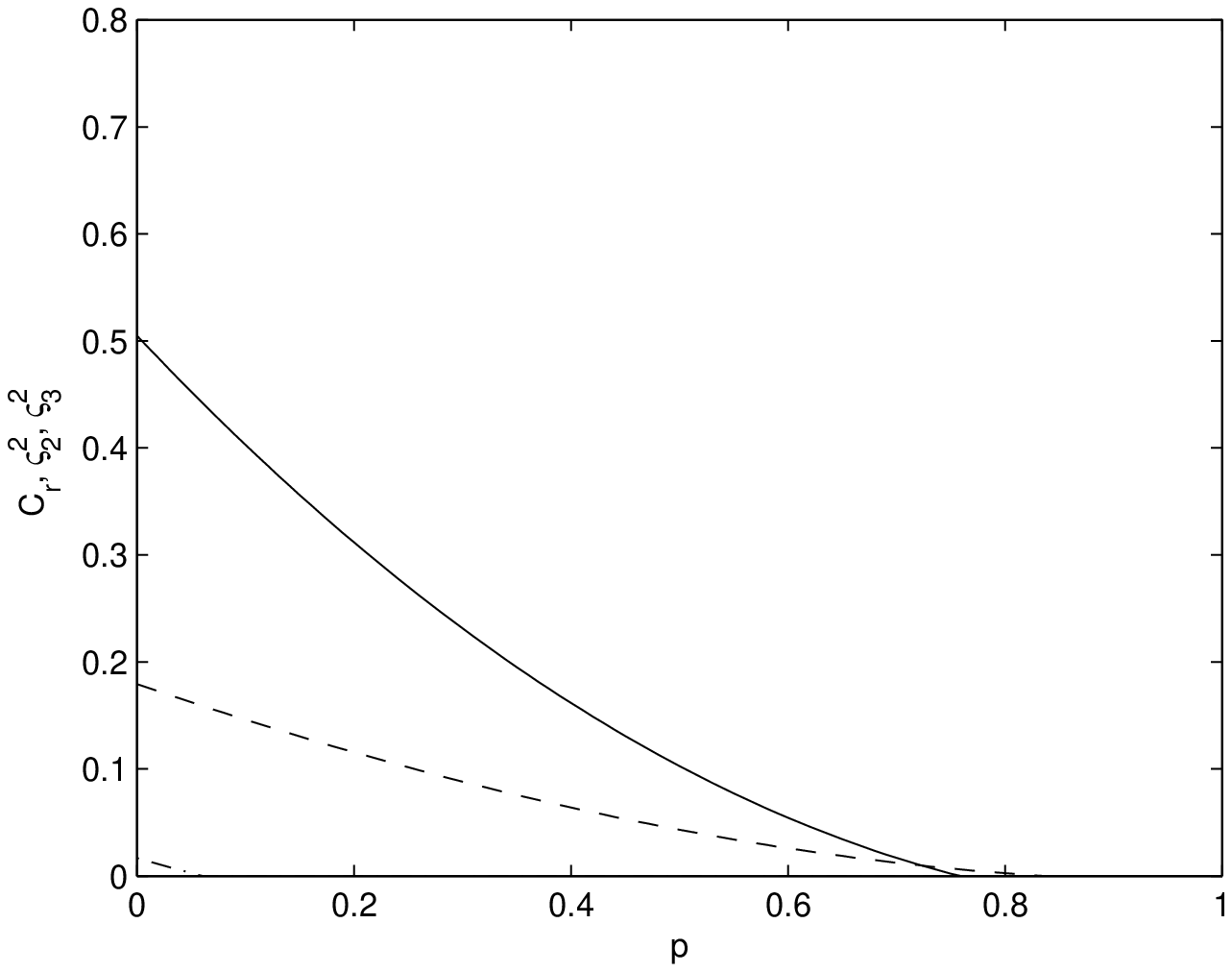}(c)
\hspace{0.01mm}
\includegraphics[width=2.7in]{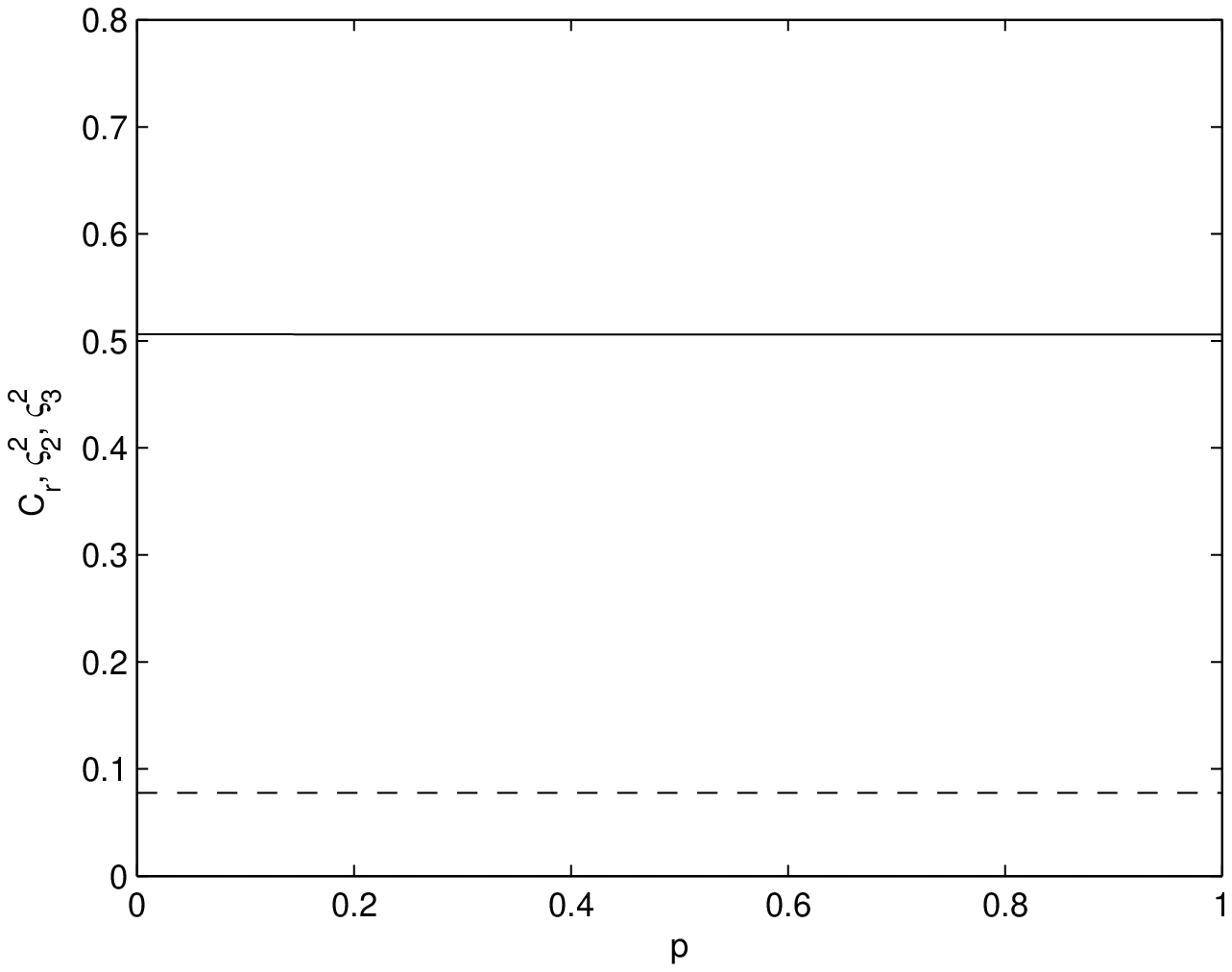}(d)
\hspace{0.01mm}
\caption{ Spin-squeezing parameters $\varsigma_2^2$ (dash-dotted line), $\varsigma_3^2$ (dashed line)
and the concurrence $C_r$ (solid line) versus the decoherence strength $p$ for the phase-damping channel with $\theta=1.8\pi$, $N=12$. (a) Without weak measurement; (b) weak measurement strength  $m=1$; (c) $m=0.5$; (d) $m=0.01$.}
  \end{figure}

\end{document}